\documentclass{aa}
\usepackage{graphics}
\usepackage{psfig}
\newcommand{\sbu}{mag arcsec$^{-2}$}

\def\h2{H{\small II}}
\def\sbs{HS 1442+4250}
\begin{document}

\title{Spectroscopic and photometric studies of low-metallicity  
star-forming dwarf galaxies. II. HS 1442+4250}

\author{N. G.\ Guseva \inst{1}
\and P.\ Papaderos \inst{2}
\and Y. I.\ Izotov \inst{1}
\and R. F. Green \inst{3}
\and K. J.\ Fricke   \inst{2}
\and T. X.\ Thuan\inst{4}
\and K.G.\ Noeske\inst{2}}
\offprints{N.G. Guseva, guseva@mao.kiev.ua}
\institute{      Main Astronomical Observatory,
                 Ukrainian National Academy of Sciences,
                 Zabolotnoho 27, Kyiv 03680,  Ukraine
\and
                 Universit\"ats--Sternwarte, Geismarlandstra\ss e 11,
                 D--37083 G\"ottingen, Germany
\and
                 National Optical Astronomy Observatory, 
                 Tucson, AZ 85726, USA
\and
                 Astronomy Department, University of Virginia, 
                 Charlottesville, VA 22903, USA
}

\date{Received \hskip 2cm; Accepted}
\titlerunning{HS 1442+4250}
\authorrunning{N.G.Guseva et al.}

\abstract{We present broad-band $V$ and $I$ imaging 
and  long-slit spectroscopy in the optical range 
$\lambda$$\lambda$3600 -- 7500\AA\
of the dwarf irregular galaxy  HS 1442+4250.
The oxygen abundance  12 + log(O/H) = 
7.63 $\pm$ 0.02 ($Z$ = $Z_\odot$/19)\thanks{12 + log(O/H)$_{\odot}$ = 8.92
(Anders \& Grevesse \cite{Anders89}).} 
in the brightest H {\sc ii} region of \sbs\
places the galaxy among the most 
metal-deficient emission-line galaxies. 
The low metallicity and blue colour ($V$ -- $I$) $\sim$ 0.4 mag of 
the low-surface-brightness (LSB) component 
make HS 1442+4250 a likely rare young dwarf galaxy candidate.
We use four methods
to estimate  the 
stellar population age in the LSB component of HS 1442+4250. 
 Different star formation histories are considered.
The equivalent widths of hydrogen H$\alpha$ and H$\beta$ emission lines,
and of hydrogen H$\gamma$ and H$\delta$ absorption lines, the
spectral energy distribution and the observed ($V-I$) colours of the LSB 
regions are 
reproduced quite well by  models with only young and intermediate-age
stellar populations. By contrast, the observational data cannot be reproduced
by a stellar population formed continuously with a constant star formation
rate in the age range from 0 to $\ga$ 2 Gyr.
While a faint old stellar population in HS 1442+4250 with an age 
$\ga$ 2 Gyr is not excluded, we find no evidence for such a population from 
the present data.
}

\maketitle

\markboth {N.G. Guseva et al.}{Spectroscopic and photometric studies of low-metallicity  
star-forming dwarf galaxies. II. HS 1442+4250}

\section {Introduction}
\label{intro}

The dwarf irregular galaxy HS 1442+4250 ($\equiv$ UGC 9497) from the 
Hamburg Survey (HS) was first
classified as an emission-line galaxy by Sanduleak \& Pesch (\cite{Sandul82}).
With the coordinates $\alpha$(J2000.0) = 14$^{\rm h}$44$^{\rm m}$12\fs1,
$\delta$(J2000.0) = +42$^\circ$37\arcmin37\arcsec\ it is situated in the 
direction of a low-density region in the galaxy spatial distribution. 
HS 1442+4250 has been studied by Tifft et al. (\cite{Tifft86}), 
Popescu et al. (\cite{Popescu96}), and Popescu \& Hopp (\cite{Popescu2000}) 
who were searching for dwarf galaxies in voids.
Popescu \& Hopp (\cite{Popescu2000}) have measured the fluxes and equivalent
widths of the emission lines in HS 1442+4250. They derived its redshift 
as $z$ = 0.0025 and an oxygen abundance 12 + log(O/H) = 7.89. 
The galaxy has also been studied spectroscopically by Kniazev et al.
(\cite{Kniazev98}) and Pustilnik et al. 
(\cite{Pustil99}),
who derived 12 + log(O/H) = 7.68 and 7.7, respectively. 

 $B$ and $R$ surface photometry of HS 1442+4250 has been  presented in Vennik 
et al. (\cite{vennik96}). This galaxy with $M_B$ = --15.2 mag
was found to be one of the bluest dwarf galaxies 
 in their sample with $(B-R)_{\rm T}$ = 0.42 mag.
This implies that the light from HS 1442+4250 is dominated by a relatively 
young stellar population.

In this paper we present  $V$ and $I$ photometric and 
 spectroscopic observations of HS 1442+4250. We study 
its properties and put constraints on the age of 
its low-surface-brightness (LSB) component.
Throughout this paper we adopt a distance of 12.4 Mpc 
for the dwarf galaxy derived 
from its redshift $z$ = 0.0025 and a Hubble constant of 
75 km s$^{-1}$ Mpc$^{-1}$, after correction for Virgocentric infall
(Kraan-Korteweg \cite{Kraan86}). At this distance 1\arcsec\ corresponds to a
linear size of 60 pc.
The structure of the paper is as follows. In Sect. \ref{obs} we describe the 
observations and data reduction. The photometric properties of HS 1442+4250 
are described in Sect. \ref{imag}. In Sect. \ref{chem} we derive
chemical abundances 
in its two brightest H {\sc ii} regions. The properties of its stellar populations 
are discussed in Sect. \ref{age}. Finally, 
Sect. \ref{conc} summarises our main conclusions.

\begin{figure*}
\begin{picture}(16,11)
\put(2,0){{\psfig{figure=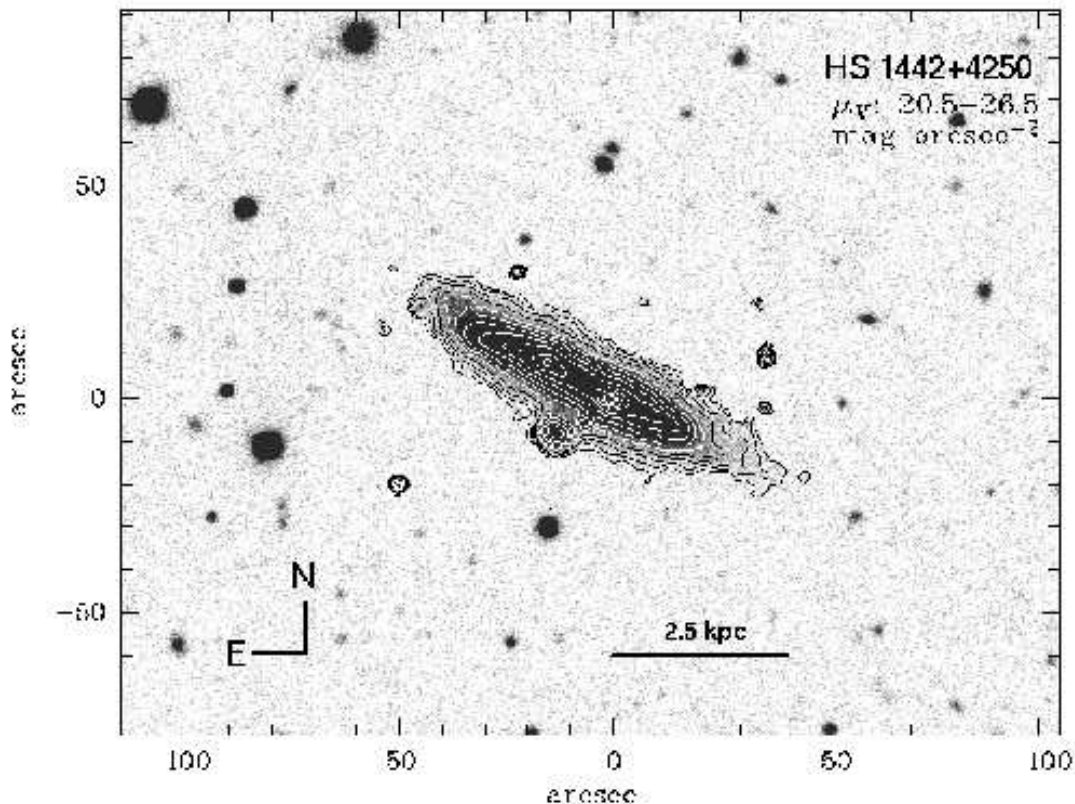,height=11cm,angle=-90,clip=}}}
\end{picture}
\caption[]{$V$ image of HS 1442+4250. The overlayed $V$ contours 
correspond to surface brightness  between 20.5 \sbu\ and 26.5 \sbu\ in
steps of 0.5 mag.} 
\label{f1}
\end{figure*}

\section{Observations and data reduction \label{obs}} 
\subsection{Photometric observations}

Broad-band $V$ and $I$ images were obtained with the Kitt Peak 
2.1m telescope\footnote{Kitt Peak National Observatory (KPNO)
is operated by the Association of Universities for Research 
in Astronomy (AURA), 
Inc., under cooperative agreement with the National 
Science Foundation (NSF).} 
on April 18, 1999 during photometric conditions. The telescope was 
equipped with a thinned Tektronix 1024 $\times$ 1024
CCD operating at a gain of 3\,e$^-$\,ADU$^{-1}$ and giving an 
instrumental scale of 0\farcs 305 pixel$^{-1}$ and a field of view of 5\arcmin. 
The total exposures of 20 min in $V$ and 40 min in $I$ were split into 
three subexposures, slightly offset with respect to each other
for removal of cosmic particle hits and bad pixels. 
The point spread function in $V$ and $I$ were respectively
1\farcs78 and 1\farcs67 FWHM.

Bias-- and flat--field exposures were taken during the same night.
The data reduction was done using IRAF\footnote{IRAF is the Image 
Reduction and Analysis Facility distributed by the National Optical Astronomy 
Observatory, which is operated by the  
AURA under cooperative agreement with the  
NSF.}. Images were calibrated by observing four different standard 
fields from Landolt (\cite{Landolt92}), each 3--4 times at different
airmasses during the same night. 
Our calibration uncertainties are estimated to be well below 0.05\,mag in all 
bands. Reduction steps included bias subtraction, 
removal of cosmic particle hits, flat--field 
correction and absolute flux calibration.

%
\begin{figure*} 
\begin{picture}(8.8,12.2)
\put(0.9,6.83){{\psfig{figure=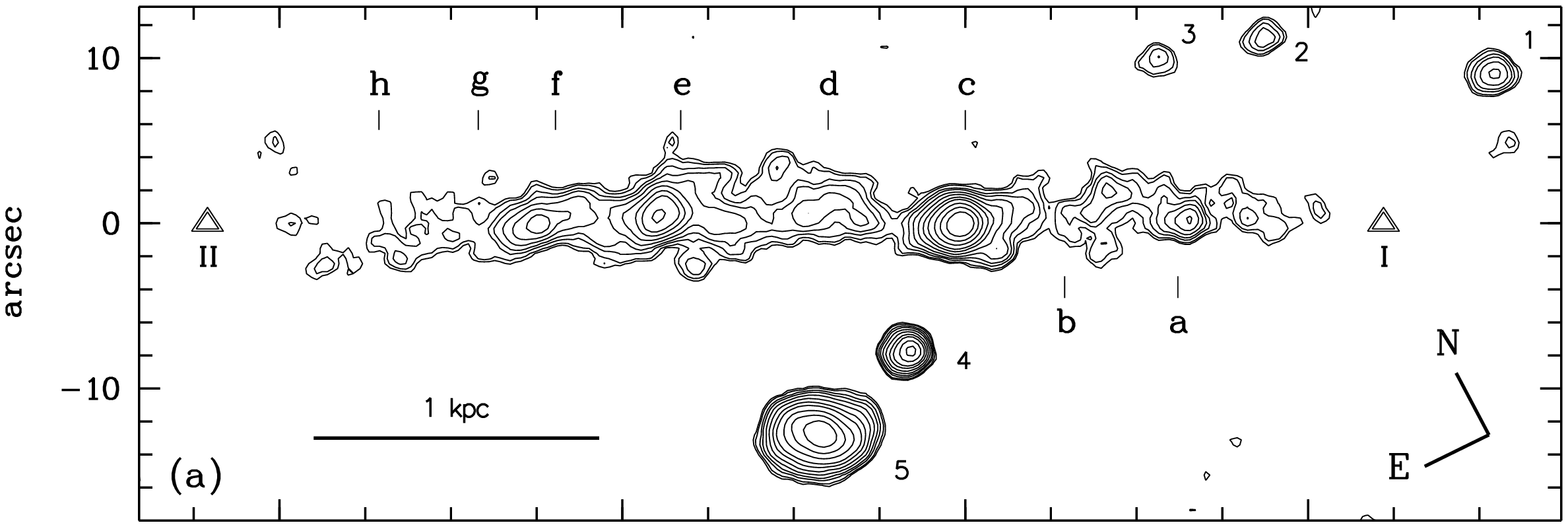,width=16cm,angle=-90.,clip=}}}
\put(1,0){{\psfig{figure=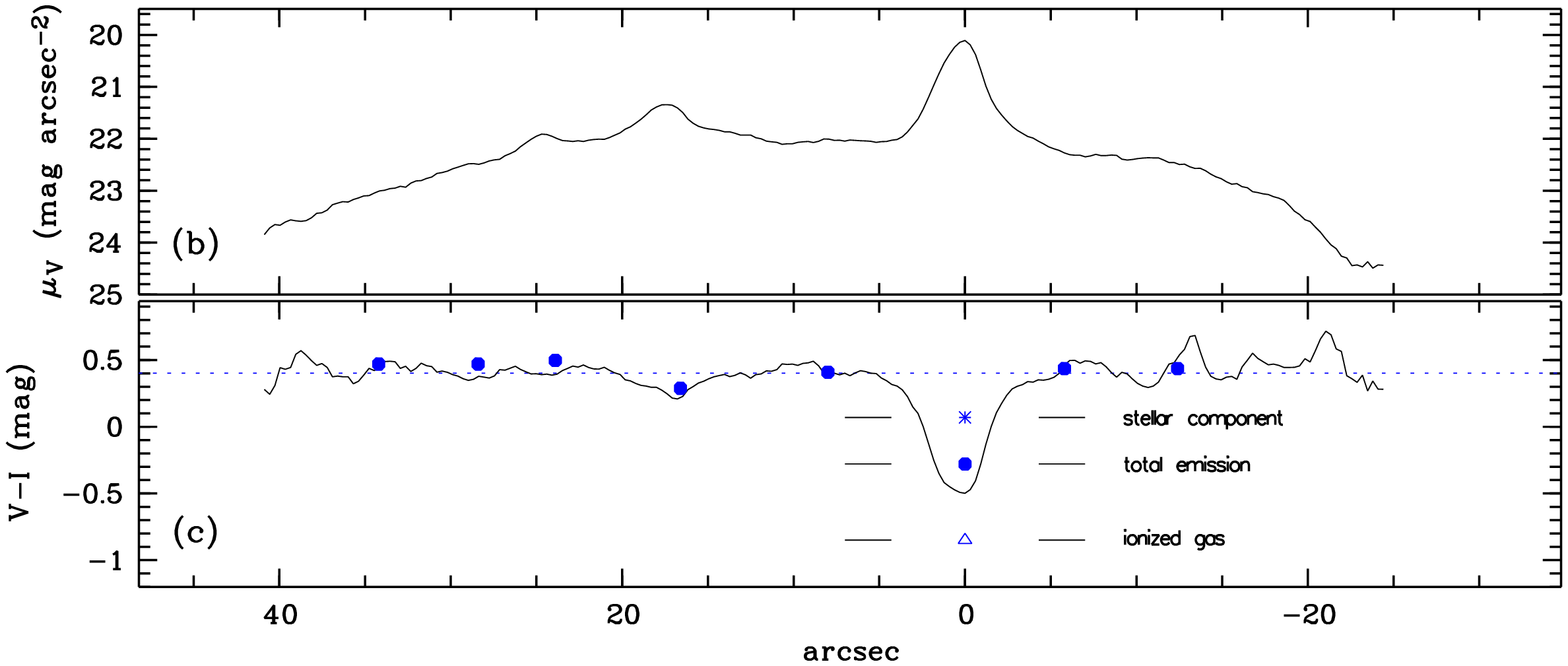,width=16cm,angle=-90.,clip=}}}
\end{picture}
\caption[]{{\bf (a)} Contrast-enhanced $I$ contour map of HS 1442+4250.
The brighter regions labeled {\it c} and {\it e} and 
the fainter regions {\it a}, {\it b}, {\it d}, {\it f}, {\it g} and 
{\it h} are arranged along the major axis of the galaxy, 
over a projected length of 
$\sim$ 1\arcmin\ (3.5 kpc). Triangles, labeled
I and II, show the positions of the outermost regions with
H$\alpha$ and H$\beta$ emission and without any absorption features.
Sources labeled 1  to 5 are significantly
redder ($0.8\leq (V-I) \la 2.0$) than the LSB component ($(V-I) \sim$ 0.4 mag) 
and most probably are not related to \sbs. 
{\bf (b)} $V$ surface brightness distribution along the major axis
of HS 1442+4250 computed within a rectangular strip of 2\arcsec\ width. 
The origin is at the location of region {\it c}.
 The six compact regions indicated in the upper panel from {\it a} to {\it f}
span a range in surface brightness between 
20 and 22.6 $V$ \sbu\ and regions {\it g} and {\it h} between 
22.3  and 23 $V$ \sbu.
{\bf (c)} $(V-I)$ colour distribution within a 2\arcsec\ strip along the
major axis of HS 1442+4250. The colour between
regions {\it a} and {\it h} is nearly constant at $\sim$ 0.4 mag (dotted line).
The bluest region ($(V-I)$ $\sim$ --0.5 mag) is coincident with 
region {\it c}. Region {\it a} presents a large 
spatial colour variation and contains a compact source which is 
much fainter on the $V$ image. This source has a maximal 
$(V-I)$ colour of 0.69 mag, being significantly redder than the
adjacent regions (see also Fig.~\ref{f3}). 
Filled 
circles show the colours predicted for a stellar population formed continuously  
with a constant star formation rate in two periods with 1) intermediate-age and 2) 
young stellar population. The filled circle for region {\it c} indicates
the colour of  a 4 Myr stellar population to which has been added the observed 
ionized gas emission in region {\it c}. The asterisk shows the colour of a 4 
Myr pure stellar population emission, while the open triangle shows the colour of ionized 
gas emission.
  }
\label{f2}
\end{figure*}


%
\subsection{Spectroscopic observations and data reduction}

 The spectroscopic observations of HS 1442+4250 were carried out on June 18, 
1999, with the Kitt Peak 4m Mayall telescope, in combination with
the Ritchey-Chr\'etien 
spectrograph and the T2KB 2048~$\times$~2048 CCD detector. The slit was
centered on the brightest star-forming region and oriented along the 
elongated body of the galaxy with position angle P.A. = 60\fdg4,
close to the direction of its major axis 
(see Fig.~\ref{f1}). The slit orientation during the observations was close
to the parallactic angle to minimize the effects of  differential refraction.
Hence, no correction was made for this effect. 
 A 2\arcsec~$\times$~300\arcsec ~slit with the KPC-10A grating in 
first order and a GG 375 order separation filter were used.
The spatial scale along the slit
was 0\farcs69 pixel$^{-1}$ and the spectral resolution $\sim$7~\AA\ (FWHM).
The spectra were obtained at an airmass of 1.44. The total 
exposure of 60 minutes was broken up into 3 subexposures. 
Two Kitt Peak spectrophotometric standard stars, 
Feige 34 and HZ 44, were observed for flux calibration.
Spectra of a He-Ne-Ar comparison lamp were obtained for wavelength calibration.

The data reduction was made with the IRAF software package. This includes 
bias subtraction, flat--field correction, cosmic-ray removal, wavelength 
calibration, night sky  subtraction, correction for atmospheric 
extinction and absolute flux calibration of the two--dimensional spectra.

One-dimensional spectra for abundance determination in the two brightest H {\sc ii} 
regions {\it c} and {\it e} (Fig.~\ref{f2}a,~\ref{f3}) were extracted 
within apertures of 2\arcsec\ $\times$ 2\farcs8. 

In addition we extracted one-dimensional spectra of six regions in the 
LSB component of the 
galaxy showing hydrogen Balmer absorption lines. The selected regions, 
labeled in Fig.~\ref{f2}a $a$, $b$, $d$, $f$, $g$ and $h$, are 
listed in Tables \ref{t:emhahb} 
and \ref{t:abshghd} together with their positions relative to the brightest region 
{\it c} and their angular extent along the slit. 
We also extracted  one-dimensional spectra of two additional outermost regions 
labeled I and II with only H$\alpha$ and H$\beta$ emission. 

\section{Imaging \label{imag}}
%
\subsection{Morphology}

\begin{figure*}
\begin{picture}(8.8,8.4)
\put(1,0){{\psfig{figure=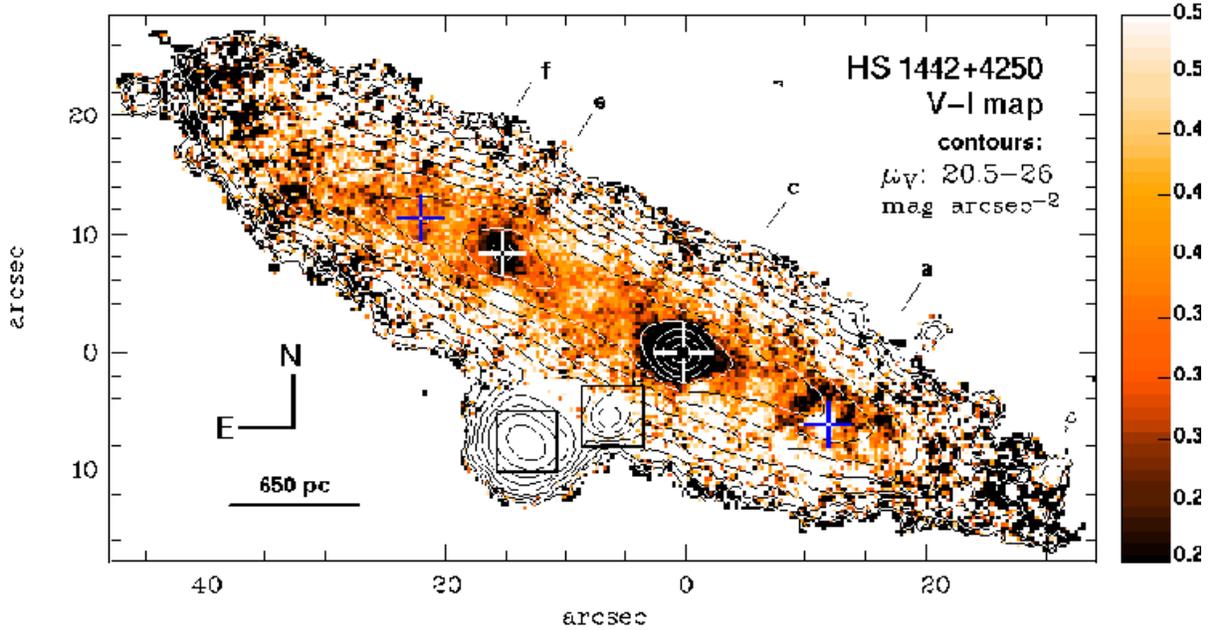,width=16cm,angle=-90,clip=}}}
\end{picture}
\caption[]{$(V-I)$ colour map of HS 1442+4250 displayed in the 
range 0.25 to 0.55 mag. The colours are corrected for foreground Galactic 
extinction ($A_V$ = 0.044 mag, $A_I$ = 0.026 mag). 
The overlayed contours are from 20.5 to 26 
$V$ mag arcsec$^{-2}$  in steps of 0.5 mag. Crosses along the major axis of
the galaxy mark the positions of regions {\it a}, {\it f} (black crosses)
and
{\it c}, {\it e} (white crosses) (see Fig. \ref{f2}a). The positions 
of the red sources 4 and 5 are marked by squares. 
The bluest region with $(V-I)$ $\sim$ --0.5 mag coincides with  
region {\it c}.} 
\label{f3}
\end{figure*}

Photometric signatures of a young to moderately evolved stellar population 
are present all along the major axis of HS 1442+4250. The $(V-I)$ colour map
(Fig. \ref{f3}) shows that the bluest region coincides with the brightest 
region {\it c}  (Fig. \ref{f2}) where ($V$ -- $I$) $\sim$ --0.5 mag. 
This region is partly resolved and has
 an effective radius of 2\farcs4 ($\approx$140 pc)
and a mean FWHM of 2\farcs5 in $V$.
Its apparent 
$V$ magnitude of 18.0 mag corresponds to an absolute $V$ magnitude 
of $\sim$ --12.5 mag. The $(V-I)$ colour 
of the second brightest region {\it e}  ($V$ = 19.65 mag)
is  $\sim$ 0.2 mag, markedly bluer 
than the local LSB component with an average $(V-I)$ colour of $\sim$ 0.4 mag.
Sources {\it c} and {\it e}
 contribute more than 80\% of the flux in excess of the 
LSB component inside the 25 $V$ mag arcsec$^{-2}$ isophote. 
Several other fainter compact sources are seen along the major axis of the galaxy
(Fig. \ref{f2}a).
Variations of the
$(V-I)$ colour on spatial scales of 2\arcsec\ -- 3\arcsec\ (Fig. \ref{f3}) 
imply that non-uniform extinction may be significant. 
  Alternatively, it may be that 
part of the reddest small regions are due to background/foreground red sources.
This is likely to be the case for the compact region {\it a}, clearly detected in both 
the $I$ image and the colour map, but marginally seen in $V$. 
Roughly 15\arcsec\ SE of region {\it c} there is a red, relatively bright 
($V$ = 18.6 mag) source labeled 5 with a mean 
$(V-I)$ $\ga$ 1.2 mag, i.e., roughly 0.6 mag redder than the LSB
component of HS 1442+4250 (see Fig. \ref{f4}b).
The surface brightness profile of this
source can be fitted by a S\'ersic profile (S\'ersic \cite{S86}) 
with a $V$ central surface brightness of $\approx$ 21.9 \sbu\ and an exponent 
$\eta$ of $\sim$ 0.7 
(see Eq.~1 in Guseva et al. \cite{Guseva2003a}). 
Most likely, this source as well as those 
labeled 1 -- 4 in Fig.~\ref{f2}a are background galaxies.

%

\begin{table*} 
\caption{\label{decomp_res}Structural properties of the starburst and  LSB
  components.}
\label{photom}
\begin{tabular}{lccccccccccc}
\hline \hline
Band & $\mu_{\rm E,0}$ & $\alpha $ &  $m_{\rm LSB}$ &  $m_{\rm LSB}^{\rm c}$ &
$P_{25}$  & $m_{P_{25}}$ & $E_{25}$ & $m_{\rm
  E_{25}}$ & $m_{\rm SBP}$ & $m_{\rm tot}$ & $r_{\rm eff}$,$r_{80}$  \\
          & mag arcsec$^{-2}$ & pc       & mag & mag & pc       &  mag             &  pc
          &  mag         &   mag & mag & pc \\
     (1) &   (2)            &   (3)     &  (4)      &   (5)            &
 (6)    &  (7)             &  (8)     &  (9)  & (10) & (11) & (12) \\
\hline
 $V$ & 20.50$\pm$0.08  & 272$\pm$~~6  & 15.23 & 15.54 & 496  & 17.65  & 1127  & 15. 65  & 
 15.37$\pm$0.02 & 15.35 & 503,822\\
 $I$ & 20.18$\pm$0.20\,  & 291$\pm$15  & 14.76 & 15.07 &
709  & 17.31  & 1290  & 15.16  & 
 14.95$\pm$0.03  & 14.93 & 556,880\\
\hline
\end{tabular}

\end{table*}

\subsection{Surface brightness and colour distribution \label{Surf}}

\begin{figure}[t]
\begin{picture}(16,12.6)
\put(0,5.3){{\psfig{figure=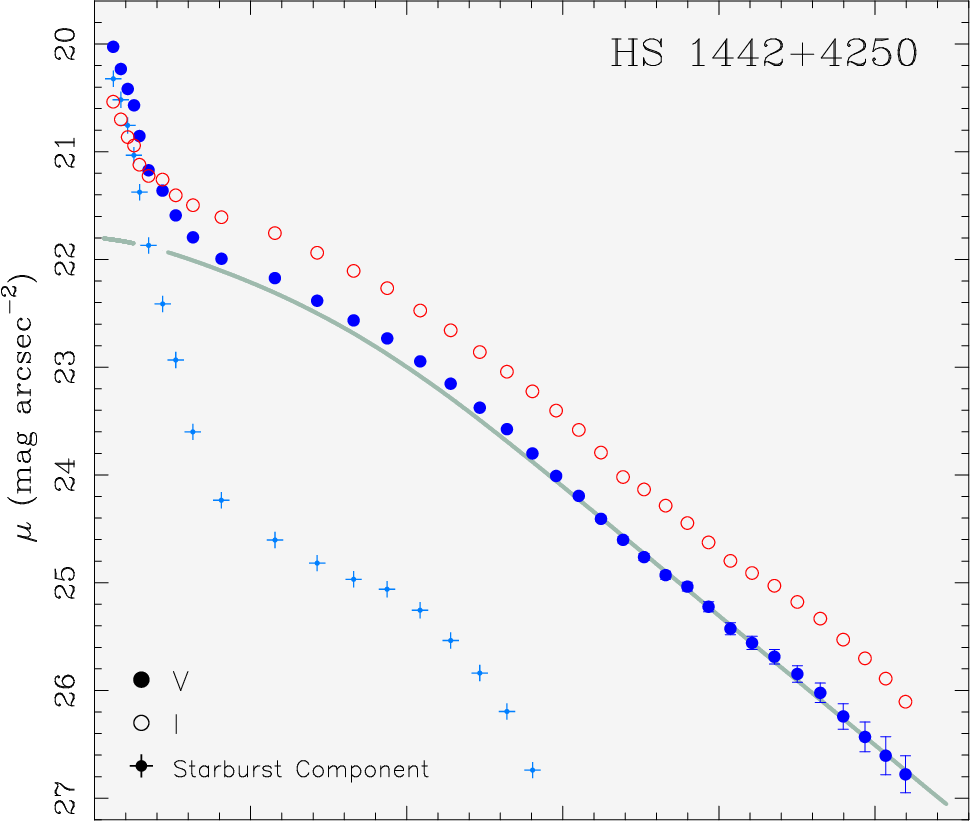,width=7.4cm,angle=-90,clip=}}}
\put(0,0){{\psfig{figure=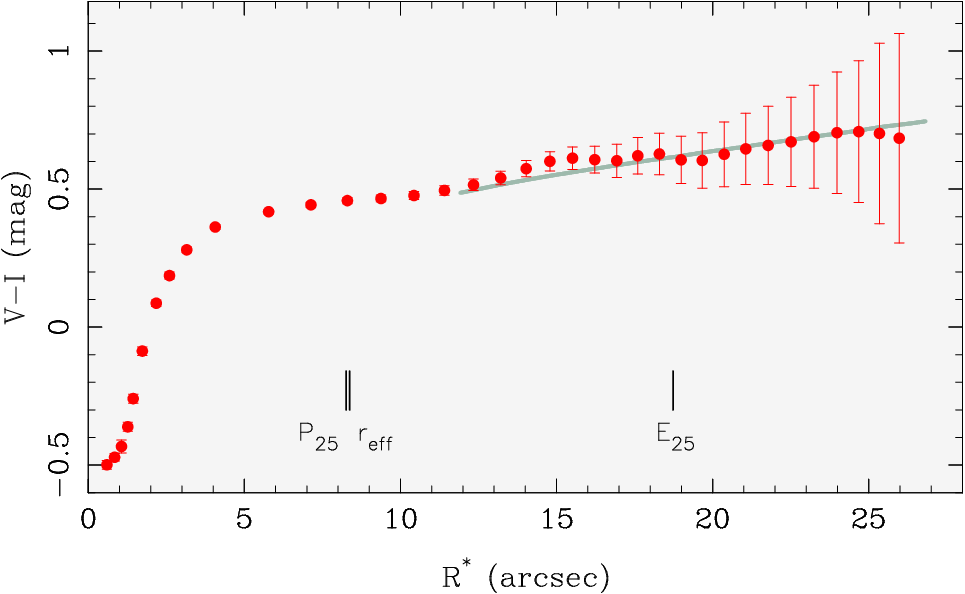,width=7.4cm,angle=-90,clip=}}}
\end{picture}
\caption[]{{\bf (a)} Surface brightness profiles (SBPs) of 
HS 1442+4250 in $V$ (filled circles) and $I$ (open circles)
derived using the method iii (Papaderos et al. \cite{Papa96a}). 
The modeled 
surface brightness distribution of the LSB component in $V$ assuming 
a modified exponential distribution by Papaderos et al. (\cite{Papa96a})
is shown by the thick-grey curve. This model implies a 
central surface brightness for the LSB component $\approx$ 1.3 mag fainter than 
the value predicted by extrapolation of the outer exponential slope of the SBPs
to $R^*=0$\arcsec. The surface brightness distribution of the light in excess 
of the LSB component (crosses) 
is due to the emission
of  star-forming regions arranged 
along the major axis of the galaxy (starburst component) (see Figs. \ref{f2} and \ref{f3}).
{\bf (b)} $(V-I)$ colour profile of HS 1442+4250 computed by 
subtraction of the SBPs displayed in the upper panel. The thick-grey curve 
shows the $(V-I)$ colour distribution  obtained by subtracting the 
adopted models  of the LSB component 
 intensity distribution in $V$ and $I$.
Both profiles suggest a slow colour increase of 0.3 mag kpc$^{-1}$ with radius
and a mean $(V-I)$ colour of 
0.63 $\pm$ 0.05 mag for $R^*\geq$ 15\arcsec.
The effective radius $r_{\rm eff}$ of the $V$ SBP
and the isophotal radii $P_{25}$ and $E_{25}$ of the starburst 
and LSB components  at 25 $V$ \sbu\ are indicated.}
\label{f4}
\end{figure}

The surface brightness profiles (SBPs) of HS 1442+4250 
in Fig. \ref{f4}a were obtained using one of the methods (method iii)
described in Papaderos et al. 
(\cite{Papa96a}) and Guseva et al. (\cite{Guseva2003a}) 
after the red compact sources 1 through 5 in 
Fig. \ref{f2}a have been fitted and subtracted from the combined $V$ and $I$ 
images. The extended sources 4 and 5 
overlap with the LSB component of \sbs\
at a surface brightness  fainter than 24.5 $V$ \sbu.
In this case, a two-dimensional 
model was fitted to their light distribution and subtracted from 
the original images. To ensure that possible residuals in the 
subtraction of sources 4 and 5 do not affect the surface photometry, 
SBPs were computed by fitting ellipses to isophotes after screening out 
the area subtended by position angles --145$^{\circ}\leq\phi\leq$ 
--95$^{\circ}$ with respect to region {\it c}.

The SBPs of HS 1442+4250 (Fig.~\ref{f4}a) are well fitted by an
exponential in the radius range 15\arcsec $\la R^* \la$ 26\arcsec,
corresponding to a surface brightness fainter than $\sim$ 24 $V$ \sbu.
In this  range we obtain a $V$ band scale length of
$\approx$ 4\farcs5, in good agreement with the value of 4\farcs2 derived 
previously in the $B$ band by Vennik et al. 
(\cite{vennik00}).

However, an inward extrapolation of the exponential fit predicts for 
radii between 5\arcsec\ and 10\arcsec\ a higher intensity 
than that observed (not shown in Fig.~\ref{f4}a). 
Therefore an exponential model flattening for small radii appears to
be necessary to adequately fit the LSB emission of \sbs.
This conclusion is in line with the results by Vennik et al. 
(\cite{vennik00}) who noted the convex shape of the SBP of HS 1442+4250
over a substantial radius range. Indeed,
for radii $\geq$ 15\arcsec, by fitting a S\'ersic (\cite{S86}) profile
to the SBPs in Fig. \ref{f4} 
we obtain an exponent $\eta$ of $\sim$ 1.2 and 1.3 in $V$ and $I$, 
respectively, slightly greater than that corresponding to a pure exponential 
distribution ($\eta$ = 1).

Alternatively, a fitting formula which reproduces an inward flattening 
of the exponential distribution has been discussed by Papaderos et al. 
(\cite{Papa96a}, their Eq. 22; see also Guseva et al. \cite{Guseva2003a}
for a detailed explanation).
Near the center, such an intensity distribution depends
on the relative central intensity depression $q=\Delta I/I_0$, where 
$I_0$ is the central intensity obtained by extrapolation
of the outer exponential slope to $R^*=0$, and on the cutoff--radius $b \alpha$
inside of which the central flattening occurs.

We find that fits to the SBPs for radii $R^*\geq 15$\arcsec\ adopting a S\'ersic
law are nearly indistinguishable from those given by Eq. 22 of Papaderos et al.  
(\cite{Papa96a}) with $b=2.1$ and $q=0.3$ (Fig. \ref{f4}a). Both 
fits predict a central surface brightness of $\sim$ 21.8 $V$ \sbu\ for the LSB 
component (Fig. \ref{f4}a), 
which is roughly 1.3 mag fainter than the value $\mu_{\rm E,0}$
predicted by extrapolation of the exponential fit 
from the outer SBP part to $R^*=0$.
The surface brightness distribution of the residual emission in excess of the 
LSB model (Fig.~\ref{f4}a, crosses)
is to be attributed to star-forming regions along the 
major axis of HS 1442+4250 (starburst component). The integrated $V$ magnitudes inside 
the 25 and 27 $V$ \sbu\ isophotes are 17.65 and 17.35 mag,
respectively. Such values
imply that the starburst component contributes $\la$ 20\% and 10\% of the 
HS 1442+4250 total
light  in the $V$ inside the respective isophotal radii. About 
one half of the total starburst component emission originates from region {\it c} 
(Fig.~\ref{f2}a,~\ref{f3}).

Table\ 1 summarizes the derived photometric quantities. 
They are not corrected for inclination and foreground Galactic extinction.
Cols.\ 2 and 3 give, 
respectively, the central surface brightness $\mu_{\rm E,0}$ and scale 
length $\alpha$ of the LSB component obtained from exponential fits to the SBPs 
for $R^*\geq 15$\arcsec\ and weighted by the photometric uncertainty of 
each point. 
For the apparent major-to-minor axis ratio of 4.1 at 26.5 $V$ \sbu\ 
and assuming a disk geometry for the LSB component
the inclination-corrected central surface brightness is $\sim$ 1.2 mag fainter
than the value quoted in col.2.
In columns 4 and 5 we show the apparent magnitudes
$m_{\rm LSB}$ of a pure exponential distribution ($q=0$) and 
$m_{\rm LSB}^{\rm c}$ of an inwards
flattening distribution (Eq.22 in Papaderos et al. 
\cite{Papa96a}) 
with the derived flattening parameters ($b,q$) = (2.1,0.3). 
Note that  both distributions  
refer to the same extrapolated central surface brightness
$\mu_{\rm E,0}$ and $\alpha$ (cols. 2 and 3).
Columns 6 through 9 list quantities obtained from profile decomposition 
whereby the intensity distribution of the LSB component was modeled by the 
modified exponential distribution proposed by
Papaderos et al. (\cite{Papa96a}). 
In cols.\ 6 and 8 we tabulate the radial extents $P_{25}$ and $E_{25}$
of the starburst and LSB components respectively, both determined 
at a surface brightness  25\ \sbu\ (see Fig. \ref{f4}b).
The respective apparent magnitudes of each component within $P_{25}$ and 
$E_{25}$ are listed in cols.\ 7 and 9.
The total magnitude of \sbs\  derived from SBP integration out to 
the outermost radius of $\approx$ 26\arcsec\ is listed in col. 10. 
The apparent magnitude of the galaxy, obtained within a polygonal aperture
after removal of sources 1 through 5 (Fig. \ref{f2}a) is given in col. 11. In 
col.\ 12 we list the effective radius $r_{\rm eff}$ and the radius $r_{80}$ 
which encircles 80\% of the galaxy's total flux.

As shown in Fig.~\ref{f4}b,  the $(V-I)$ colour of HS 1442+4250 
beyond the radius $P_{25}$
increases with an average gradient of 0.24 mag kpc$^{-1}$. 
The observed gradual colour increase 
is in agreement with the linear slope 
derived by subtracting the modeled distributions 
for the LSB component (thick-gray lines in Fig. \ref{f4}b).
From the colour profile we derive
a mean ($V-I$) = 0.63 $\pm$ 0.05 mag 
in the radius range 
15\arcsec $\leq R^* \leq$ 26\arcsec.
The ($V-I$) colour at  $R^* \ge$ 18\arcsec\  is uncertain, and 
it is not clear whether the colour gradient is attributable to \sbs, or
caused by the background sources and residuals after their subtraction.

\section{Chemical abundances \label{chem}}

\begin{figure*}[hbtp]
   \hspace*{2.5cm}\psfig{figure=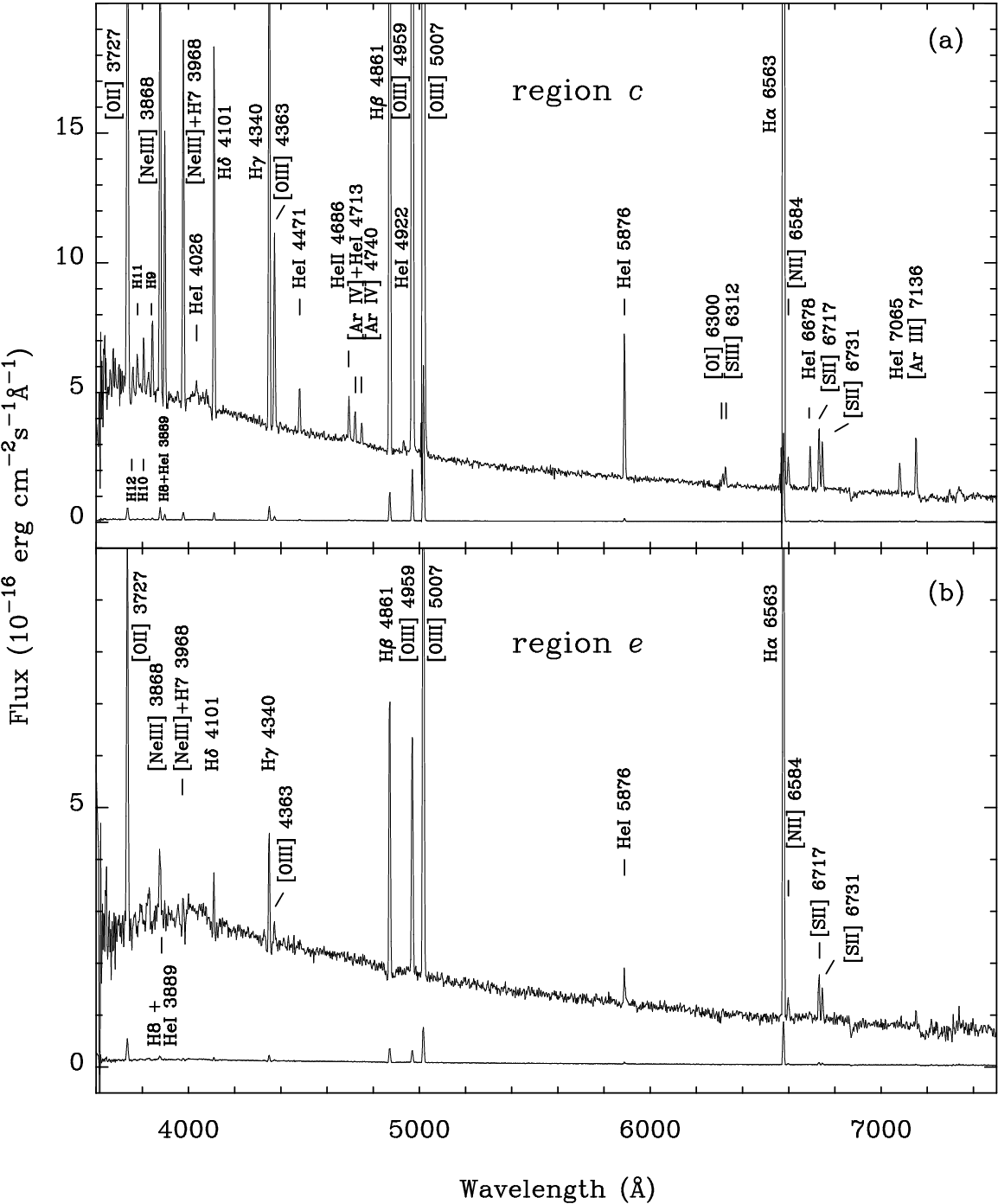,angle=0,width=14cm}
    \caption{The spectra of regions {\it c} and {\it e}. 
     The lower spectra in (a) and (b) are the observed spectra downscaled by 
     factors of 50 and 20, respectively.
      }
    \label{fig:brightsp}
\end{figure*}

In this Section we analyze the element abundances in 
the two brightest H {\sc ii} regions {\it c} and {\it e}. 
Their spectra, shown in Fig. \ref{fig:brightsp},  are characterised
by strong nebular emission lines. 

The emission line fluxes were measured by Gaussian profile 
fitting. The errors of the line fluxes include the errors
in the fitting of profiles and those in the placement of the continuum. 
They have been propagated in the calculations of the elemental abundance errors.
The spectra were corrected for interstellar extinction with
the extinction coefficient $C$(H$\beta$) derived from the 
hydrogen Balmer decrement using the equations from 
Izotov et al. (\cite{ITL94}) and the theoretical hydrogen emission line 
flux ratios from Brocklehurst (\cite{Brocklehurst71}).
The emission line fluxes relative to the 
H$\beta$ emission line flux, both observed ($F$($\lambda$)) and corrected for extinction and underlying
stellar hydrogen absorption ($I$($\lambda$)), the equivalent widths $EW$ of the emission 
lines, the observed fluxes of the H$\beta$ emission line, and the equivalent 
widths of the hydrogen absorption lines for regions {\it c} and {\it e} 
are listed in Table \ref{t:Intens}.

Our measured fluxes of the emission lines in region {\it c} are in fair 
agreement with those from Popescu \& Hopp (\cite{Popescu2000}) except for 
the [O {\sc ii}]$\lambda$3727\AA\ emission line which in our case is $\sim$ 2 
times weaker. The agreement between our measured emission line fluxes 
and those by Kniazev et al. (\cite{Kniazev98}) and 
Pustilnik et al. (\cite{Pustil99}) is much better.

Note that the flux of the He {\sc ii} $\lambda$4686 nebular emission line 
in region {\it c} is relatively high ($\sim$ 3\% of the H$\beta$ flux) 
(Table \ref{t:Intens}) implying that hard radiation beyond $\lambda$228\AA\
is intense. Such a property is not unusual for hot supergiant H {\sc ii} regions
in low-metallicity dwarf galaxies. Strong He {\sc ii} $\lambda$4686 nebular emission
has also been detected in I Zw 18 ($Z_\odot$/50, e.g., Izotov et al. 
\cite{ICFGGT99}), SBS 0335--052 ($Z_\odot$/40, Izotov et al. \cite{ILCFGK97}, 
\cite{ICFGGT99}), Tol 1214--277 ($Z_\odot$/23, 
e.g., Fricke et al. \cite{Fricke00}), 
SBS 1415+437 ($Z_\odot$/21, Thuan et al. \cite{til95}, \cite{ti99}) and some
other low-metallicity galaxies (Guseva et al. \cite{Guseva2000}).

Only a few strongest emission lines are detected in the spectra of the 
two outermost regions I and II. 
The observed fluxes and $EW$s of the H$\alpha$ and H$\beta$ 
emission lines in these regions 
are listed in Table~\ref{t:emhahb}. Since the continuum in the
red part of the spectra in those regions is weak, the $EW$(H$\alpha$) are very 
uncertain. The extinction coefficient $C$(H$\beta$) for regions I and II is set
to 0.

The photoionization model used to convert line fluxes into abundances
is the same as described in Guseva et al. (\cite{Guseva2003a}).
The ionic and heavy element abundances for regions {\it c} and {\it e}  
together with electron temperatures, 
electron number densities and adopted ICFs are given in Table \ref{t:Chem}.

\begin{table*}[tbh]
\caption{Observed ($F$($\lambda$)) and corrected  
($I$($\lambda$)) fluxes and equivalent widths ($EW$) of emission lines
 in  regions {\it c} and {\it e}.}
\label{t:Intens}
\begin{tabular}{lccrcccc} \hline \hline
  &\multicolumn{3}{c}{region {\it c}}&&\multicolumn{3}{c}{region {\it e}} \\ \cline{2-4} \cline{6-8}
$\lambda_{0}$(\AA) Ion                  &$F$($\lambda$)/$F$(H$\beta$)
&$I$($\lambda$)/$I$(H$\beta$)&$EW$(\AA)&&$F$($\lambda$)/$F$(H$\beta$)&$I$($\lambda$)/$I$(H$\beta$) 
&$EW$(\AA)  \\ \hline
3727\ [O {\sc ii}]       & 0.496 $\pm$0.008 & 0.535 $\pm$0.009 &38.3 $\pm$0.8 && 1.477 $\pm$0.089 &1.376 $\pm$0.094 &24.4 $\pm$1.8 \\
3750\ H12                & 0.020 $\pm$0.004 & 0.027 $\pm$0.006 & 1.6 $\pm$0.4 &&       ...        &       ...       &        ...     \\
3771\ H11                & 0.028 $\pm$0.004 & 0.035 $\pm$0.006 & 2.1 $\pm$0.4 &&       ...        &       ...       &        ...     \\
3798\ H10                & 0.034 $\pm$0.003 & 0.042 $\pm$0.006 & 2.6 $\pm$0.4 &&       ...        &       ...       &        ...     \\
3835\ H9                 & 0.056 $\pm$0.004 & 0.065 $\pm$0.006 & 4.4 $\pm$0.4 &&       ...        &       ...       &        ...     \\
3868\ [Ne {\sc iii}]     & 0.405 $\pm$0.007 & 0.432 $\pm$0.008 &32.1 $\pm$0.7 && 0.260 $\pm$0.048 &0.242 $\pm$0.048 &4.0 $\pm$1.4  \\
3889\ H8\ +\ He {\sc i}  & 0.176 $\pm$0.005 & 0.193 $\pm$0.007 &15.0 $\pm$0.6 && 0.070 $\pm$0.033 &0.171 $\pm$0.089 &1.2 $\pm$1.0  \\
3968\ [Ne {\sc iii}] + H7& 0.280 $\pm$0.006 & 0.302 $\pm$0.007 &24.1 $\pm$0.7 && 0.144 $\pm$0.034 &0.237 $\pm$0.078 &2.5 $\pm$1.0  \\
4026\ He {\sc i}         & 0.015 $\pm$0.004 & 0.016 $\pm$0.004 & 2.0 $\pm$0.5 &&       ...        &       ...       &        ...     \\
4101\ H$\delta$          & 0.260 $\pm$0.006 & 0.277 $\pm$0.007 &24.9 $\pm$0.7 && 0.174 $\pm$0.032 &0.261 $\pm$0.065 &3.1 $\pm$1.0  \\
4340\ H$\gamma$          & 0.467 $\pm$0.008 & 0.486 $\pm$0.008 &50.0 $\pm$1.0 && 0.374 $\pm$0.039 &0.438 $\pm$0.056 &7.4 $\pm$1.4  \\
4363\ [O {\sc iii}]      & 0.134 $\pm$0.004 & 0.138 $\pm$0.004 &14.2 $\pm$0.7 && 0.058 $\pm$0.028 &0.055 $\pm$0.028 &1.3 $\pm$1.2  \\
4388\ He {\sc i}         & 0.005 $\pm$0.002 & 0.005 $\pm$0.002 & 0.6 $\pm$0.4 &&       ...        &       ...       &        ...     \\
4471\ He {\sc i}         & 0.034 $\pm$0.003 & 0.035 $\pm$0.003 & 3.8 $\pm$0.5 &&       ...        &       ...       &        ...     \\
4686\ He {\sc ii}    & 0.030 $\pm$0.003 & 0.030 $\pm$0.003 & 3.7 $\pm$0.6 &&       ...        &       ...       &        ...     \\
4713\ [Ar {\sc iv}] + He {\sc i} & 0.022 $\pm$0.003 & 0.022 $\pm$0.003 & 2.7 $\pm$0.6 &&       ...        &       ...       &        ...     \\
4740\ [Ar {\sc iv}]      & 0.014 $\pm$0.002 & 0.014 $\pm$0.002 & 1.8 $\pm$0.5 &&       ...        &       ...       &        ...     \\
4861\ H$\beta$           & 1.000 $\pm$0.012 & 1.000 $\pm$0.012 &147.1 $\pm$1.9&& 1.000 $\pm$0.063 &1.000 $\pm$0.070 &25.7 $\pm$2.5 \\
4922\ He {\sc i}         & 0.010 $\pm$0.003 & 0.010 $\pm$0.003 & 1.4 $\pm$0.6 &&       ...        &       ...       &        ...    \\ 
4959\ [O {\sc iii}]      & 1.769 $\pm$0.019 & 1.754 $\pm$0.018 &250.0 $\pm$2.6&& 0.831 $\pm$0.057 &0.774 $\pm$0.057 &20.1 $\pm$2.4 \\
5007\ [O {\sc iii}]      & 5.032 $\pm$0.046 & 4.976 $\pm$0.046 &782.3 $\pm$4.3&&2.489 $\pm$0.130 &2.317 $\pm$0.130  &62.5 $\pm$6.9 \\
5876\ He {\sc i}         & 0.099 $\pm$0.003 & 0.094 $\pm$0.003 &22.7 $\pm$1.3 && 0.100 $\pm$0.026 &0.093 $\pm$0.026 & 3.6 $\pm$2.0 \\
6300\ [O {\sc i}]        & 0.009 $\pm$0.002 & 0.008 $\pm$0.002 & 2.4 $\pm$0.8 &&       ...        &       ...       &        ...     \\
6312\ [S {\sc iii}]      & 0.016 $\pm$0.002 & 0.014 $\pm$0.002 & 4.4 $\pm$0.9 &&       ...        &       ...       &        ...     \\
6563\ H$\alpha$          & 3.012 $\pm$0.028 &2.763 $\pm$0.029 &881.8 $\pm$6.8 &&2.900 $\pm$0.147 &2.738 $\pm$0.162 &133.8 $\pm$6.9 \\
6584\ [N {\sc ii}]       & 0.025 $\pm$0.002 & 0.023 $\pm$0.002 & 7.5 $\pm$1.0 && 0.070 $\pm$0.019 &0.066 $\pm$0.019 & 3.3 $\pm$2.0 \\
6678\ He {\sc i}         & 0.028 $\pm$0.002 & 0.026 $\pm$0.002 & 8.6 $\pm$1.1 &&       ...        &       ...       &        ...     \\
6717\ [S {\sc ii}]       & 0.046 $\pm$0.002 & 0.042 $\pm$0.002 &13.9 $\pm$1.2 && 0.147 $\pm$0.024 &0.137 $\pm$0.025 & 7.0 $\pm$2.5 \\
6731\ [S {\sc ii}]       & 0.032 $\pm$0.002 & 0.029 $\pm$0.002 & 9.7 $\pm$1.1 && 0.095 $\pm$0.020 &0.088 $\pm$0.020 & 4.6 $\pm$2.1 \\
7065\ He {\sc i}         & 0.023 $\pm$0.002 & 0.020 $\pm$0.002 & 7.8 $\pm$1.1 &&       ...        &       ...       &        ...     \\
7136\ [Ar {\sc iii}]     & 0.043 $\pm$0.002 & 0.038 $\pm$0.002 &15.8 $\pm$1.4 &&       ...        &       ...       &        ...     \\
                     & & & & & & & \\
$C$(H$\beta$)\ dex             &\multicolumn {3}{c}{0.110$\pm$0.013} &&\multicolumn {3}{c}{0.000$\pm$0.063} \\
$F$(H$\beta$)$^{\rm a}$              &\multicolumn {3}{c}{3.85$\pm$0.03}   &&\multicolumn {3}{c}{0.44$\pm$0.02} \\
$EW$(abs)~\AA                  &\multicolumn {3}{c}{0.4$\pm$0.4}  &&\multicolumn {3}{c}{1.45$\pm$1.0} \\
\hline
\end{tabular}

$^{\rm a}$in units 10$^{-14}$\ erg\ s$^{-1}$cm$^{-2}$.
\end{table*}


\subsection{Heavy element abundances}

The oxygen abundances in regions {\it c} and {\it e} are 12 + log(O/H) = 
7.63 $\pm$ 0.02 ($Z$ $\sim$ $Z_\odot$/19) and 7.54 $\pm$ 0.20 
($Z$ $\sim$ $Z_\odot$/24), respectively. These two determinations agree within 
the errors. 
However, the oxygen abundance
in region {\it e} may have been underestimated. 
This is because the weak 
[O {\sc iii}]$\lambda$4363 emission line 
[$I$($\lambda$4363)/$I$(H$\beta$)=0.055 (Table~\ref{t:Intens})]
may be enhanced by shocks. This effect is likely lower in region {\it c} 
where the [O {\sc iii}]$\lambda$4363 emission line is stronger.
For comparison, Popescu \& Hopp (\cite{Popescu2000}) 
obtained for HS 1442+4250 
12 + log(O/H) = 7.89, while Kniazev et al. (\cite{Kniazev98}) and
Pustilnik et al. (\cite{Pustil99}) derived 
12 + log(O/H) = 7.68 and 7.7, respectively. The former value is 
significally larger than ours due to the higher [O {\sc ii}]$\lambda$3727\AA\ 
and 
lower [O {\sc iii}]$\lambda$4363\AA\ fluxes measured by Popescu \& 
Hopp (\cite{Popescu2000}).

The neon-to-oxygen abundance ratio  log Ne/O = --0.75 
for region {\it c} is 
in good agreement with the mean ratio derived from previous studies of
BCDs (e.g., Izotov \& Thuan \cite{IT99}).
The nitrogen-to-oxygen abundance ratio log N/O = --1.44 is $\sim$ 0.15 dex  
higher than the N/O ratios obtained
by Thuan et al. (\cite{til95}) and Izotov \& Thuan (\cite{IT99}) for the very
metal-deficient BCDs with $Z$ $\la$ $Z_\odot$/20. 

\begin{table}[tbh]
\caption{Element abundances in regions {\it c} and {\it e}.} 
\label{t:Chem}
\begin{tabular}{lccc} \hline \hline
Value                               & region {\it c}      && region {\it e}  \\ \hline
$T_{\rm e}$(O {\sc iii})(K)               &17810$\pm$320   && 16410$\pm$4100 \\
$T_{\rm e}$(O {\sc ii})(K)                &14960$\pm$250   && 14390$\pm$3420 \\
$T_{\rm e}$(S {\sc iii})(K)               &16480$\pm$260   && 15320$\pm$3400 \\
$N_{\rm e}$(S {\sc ii})(cm$^{-3}$)        & 10$\pm$10  &&  10$\pm$10   \\ \\
O$^+$/H$^+$($\times$10$^5$)         &0.467$\pm$0.022 && 1.372$\pm$0.890\\
O$^{+2}$/H$^+$($\times$10$^5$)      &3.685$\pm$0.159 && 2.067$\pm$1.279\\
O$^{+3}$/H$^+$($\times$10$^6$)      &1.464$\pm$0.194 &&      ...       \\
O/H($\times$10$^5$)                 &4.299$\pm$0.188 && 3.439$\pm$1.558\\
12 + log(O/H)                       &7.63$\pm$0.02   && 7.54$\pm$0.20\\ \\
N$^{+}$/H$^+$($\times$10$^7$)       &1.702$\pm$0.128  && 5.361$\pm$2.585\\
ICF(N)$^{\rm a}$                    &9.20\,~~~~~~~~~  && 2.51\,~~~~~~~~~~\\
log(N/O)                            &--1.438$\pm$0.038~~&&--1.408$\pm$0.287~~\\ \\
Ne$^{+2}$/H$^+$($\times$10$^5$)     &0.662$\pm$0.035&& 0.467$\pm$0.300\\
ICF(Ne)$^{\rm a}$                   &1.17\,~~~~~~~~~&&1.66\,~~~~~~~~~~\\
log(Ne/O)                           &--0.746$\pm$0.030~~&&--0.646$\pm$0.342~~\\ \\
S$^+$/H$^+$($\times$10$^7$)         &0.700$\pm$0.036  &&       ...      \\
S$^{+2}$/H$^+$($\times$10$^7$)      &5.593$\pm$0.725  &&       ...      \\
ICF(S)$^{\rm a}$                    &2.30\,~~~~~~~~~&&        ...      \\
log(S/O)                            &--1.473$\pm$0.054~~&&     ...      \\ \\
Ar$^{+2}$/H$^+$($\times$10$^7$)      &1.185$\pm$0.070  &&      ...      \\
Ar$^{+3}$/H$^+$($\times$10$^7$)      &1.420$\pm$0.252  &&      ...      \\
ICF(Ar)$^{\rm a}$                    &1.01\,~~~~~~~~~&&       ...      \\
log(Ar/O)                            &--2.213$\pm$0.048~~&&    ...      \\ \\
He$^+$/H$^+$($\lambda$4471)         &0.075$\pm$0.007&&       ...      \\
He$^+$/H$^+$($\lambda$5876)         &0.078$\pm$0.003&&       ...      \\
He$^+$/H$^+$($\lambda$6678)         &0.077$\pm$0.006&&       ...      \\
He$^+$/H$^+$ (mean)                 &0.078$\pm$0.002&&       ...       \\
He$^{+2}$/H$^+$($\lambda$4686)      &0.003$\pm$0.000&&      ...        \\
He/H ($\lambda$4471)                &0.078$\pm$0.007&&       ...       \\
He/H ($\lambda$5876)                &0.081$\pm$0.003&&       ...       \\
He/H ($\lambda$6678)                &0.080$\pm$0.006&&       ...       \\
He/H (mean)                         &0.080$\pm$0.002&&       ...       \\
$Y$ ($\lambda$4471)                 &0.237$\pm$0.022&&       ...     \\
$Y$ ($\lambda$5876)                 &0.244$\pm$0.009&&       ...     \\
$Y$ ($\lambda$6678)                 &0.242$\pm$0.018&&       ...     \\
$Y$ (mean)                          &0.243$\pm$0.008&&       ...     \\ \hline
\end{tabular}

$^{\rm a}$ICF is the ionization correction factor.
\end{table}

\subsection{He abundances}

The He {\sc i} emission lines in region {\it c} are strong, allowing a reliable
determination of the He abundance. Five He {\sc i} $\lambda$3889, 
$\lambda$4471, $\lambda$5876, $\lambda$6678, $\lambda$7065 emission lines are 
used to correct their fluxes self-consistently for collisional and fluorescent 
enhancement (see the method description in Izotov et al. \cite{ITL94}).
Then the He$^+$ abundance is derived from the corrected He {\sc i} 
$\lambda$4471, $\lambda$5876, $\lambda$6678 emission line fluxes. Some 
fraction of He ($\sim$ 3.8\%) is in the He$^{+2}$ form as derived from the 
He {\sc ii} $\lambda$4686 emission line flux. 
The total helium abundance is derived
as the sum of the He$^+$ and He$^{+2}$ abundances and is shown
in Table \ref{t:Chem} for each of the three He {\sc i} lines used for the 
abundance determination. The mean value of the He mass fraction $Y$ = 
0.243 $\pm$ 0.008 obtained for region {\it c} (see Table \ref{t:Chem}), is 
consistent with the primordial $^4$He mass fraction $Y_{\rm p}$ = 0.244 
$\pm$ 0.002, derived by extrapolating the $Y$ vs O/H linear 
regression to O/H = 0 (Izotov \& Thuan \cite{IT98}), and with $Y_{\rm p}$ = 
0.245 $\pm$ 0.002 derived for the two
most metal-deficient BCDs, known, I Zw 18 and SBS 0335--052 (Izotov et al.
\cite{ICFGGT99}). However, the He abundance in
region {\it c} is likely  slightly underestimated because of
underlying stellar He {\sc i} absorption. Due to the relatively
small equivalent widths of the He {\sc i} emission lines (Table \ref{t:Intens})
the effect of the underlying absorption may lead to an underestimate of the 
He$^+$ abundance by 1\% -- 2\%
in the case of the strongest He {\sc i} $\lambda$5876 line. This effect is 
larger for the weaker He {\sc i} $\lambda$4471 and $\lambda$6678 emission lines.
In region {\it e} only weak He {\sc i} $\lambda$3889 and $\lambda$5876 emission
lines are observed. Therefore, an accurate He abundance determination 
is not possible in that region.

\section{The stellar populations \label{age}}

The low metallicity of HS 1442+4250 suggests that it may be
a young nearby dwarf galaxy (Izotov $\&$ Thuan \cite{IT99}). 
However, the morphological properties
of the galaxy, such as the presence of multiple star-forming regions 
on the one hand, and a redder extended diffuse stellar component
on the other hand, suggest that stars in HS 1442+4250
have been formed during several episodes of star formation. 
We study in this section the properties of the stellar populations
in the extended  LSB component and discuss the evolutionary status
of HS 1442+4250.

For this we use four different methods, three of them based on
the spectroscopic data and the fourth one using photometric
data. Two of these methods are based on the time evolution of equivalent widths of 
hydrogen Balmer line emission and absorption which are detected in almost 
all regions along the slit. The  other two methods are based on the comparison of 
observed with theoretical spectral energy distributions (SED) and broad-band
colours.
All the aforementioned methods
are sensitive to the star formation history.
The first two methods are extinction-insensitive, while the others 
depend on both the properties of the stellar population and 
the interstellar extinction. 
The use of all four methods allows us to put constraints
on the star formation history and properties of the stellar population as well
as to estimate the interstellar extinction. These methods are described
in detail in Guseva et al. (\cite{Guseva2001,Guseva2003a}).

The observed $(V-I)$ colour of the brightest region {\it c} is very blue
$\sim$ --0.5 (Fig. \ref{f2}c), and  cannot be reproduced  by a
stellar population of any age. This is because of the large contribution of very
strong oxygen and hydrogen emission lines,  as evidenced 
by their large equivalent widths (Table \ref{t:Intens}), and gaseous continuum.
Gaseous emission is also an important contributor to the total light of regions
I and II. However, in other regions the equivalent widths of the emission 
lines are small (Table \ref{t:emhahb}). Therefore, for those regions
we will not take into account ionized gas emission.
 
\subsection{Age from Balmer nebular emission lines\label{under_1}}

The equivalent widths of hydrogen nebular emission lines are 
usually used as an age indicator of star-forming regions, in which O and 
early B stars are still present. The equivalent widths of these lines 
for an instantaneous burst decrease sharply after
$\sim$ 10 Myr (e.g., Schaerer \& Vacca \cite{SV98}). However, in the case of 
continuous star formation this method can be succesfully used for age 
determination in the 0 -- 10 Gyr range (Guseva et al. \cite{Guseva2001}).

For the age determination we use the two brightest hydrogen 
emission lines, H$\alpha$ and H$\beta$. Their fluxes and equivalent 
widths, measured in the spectra 
of the extracted LSB regions  are shown with 
errors in Table~\ref{t:emhahb}.

Because the H$\beta$ emission line is narrower than the absorption line in
these regions and does not fill the absorption component, its flux was measured
using the continuum level at the bottom of the absorption line.
This level has been chosen by visually interpolating from the absorption line
wings to the center of the line.

The extinction coefficients $C$(H$\beta$) in regions 
with detectable H$\alpha$ and H$\beta$ emission lines
are derived 
from the observed H$\alpha$/H$\beta$ flux ratios.
The theoretical recombination H$\alpha$/H$\beta$ flux ratio of
2.8 is adopted, which is typical for low-metallicity H {\sc ii} regions.
No correction for the absorption line equivalent widths has been made.
The extinction coefficients $C$(H$\beta$) 
are shown in Table \ref{t:emhahb}. 
Note, that $C$(H$\beta$) for region $f$ is uncertain because of the low 
equivalent width of the H$\beta$ emission line which is comparable to
the equivalent width of the absorption line.


\begin{table*}[tbh]
\caption{Fluxes, equivalent widths of H$\alpha$ and H$\beta$  
emission lines and the extinction coefficient $C$(H$\beta$) 
in LSB regions.}
\label{t:emhahb}
\begin{tabular}{lcccrrrrr} \hline \hline
Region &\multicolumn{1}{c}{Distance$^{\rm a}$}& \multicolumn{1}{c}{Aperture$^{\rm b}$}
& {} & \multicolumn{1}{c}{$F$(H$\alpha$)$^{\rm c}$}
  &\multicolumn{1}{c}{$EW$(H$\alpha$)$^{\rm d}$} 
&\multicolumn{1}{c}{$F$(H$\beta$)$^{\rm c}$}  &\multicolumn{1}{c}{$EW$(H$\beta$)$^{\rm d}$} 
&\multicolumn{1}{c}{$C$(H$\beta$)}
   \\ \hline

I    &--24.4~~  &2.0$\times$4.1 & & 8.0  $\pm$0.5 & 1200.0 $\pm$2.5 & 3.0  $\pm$0.5 & 32.8 $\pm$0.8   & 0.0~~~~ \\ 
{\it a}  &--12.4~~  &2.0$\times$7.6 & & 50.4 $\pm$0.6 & 78.9   $\pm$1.8 & 17.9 $\pm$0.5 & 14.6 $\pm$0.7   & 0.00   $\pm$0.07   \\
{\it b}  &--5.8\,   &2.0$\times$4.1 & & 22.0 $\pm$0.6 & 36.5   $\pm$0.8 & 7.2  $\pm$0.6 & 7.3  $\pm$0.5   & 0.12  $\pm$0.07   \\
{\it d}  & ~\,8.0   &2.0$\times$5.5 & & 53.9 $\pm$0.3 & 52.4   $\pm$2.5 & 17.1 $\pm$0.6 & 9.9  $\pm$0.5   & 0.15  $\pm$0.07   \\
{\it f}  &23.9   &2.0$\times$4.1 & & 13.6 $\pm$0.6 & 16.8   $\pm$1.8 & 2.8:$\pm$0.5 & 3.1  $\pm$0.7  & 0.31:$\pm$0.07   \\
{\it g}  &28.4   &2.0$\times$4.8 & & 17.4 $\pm$0.6 & 31.7   $\pm$0.8 & ...~~~~       & ...~~~~         & ...~~~~~          \\
{\it h}  &34.2   &2.0$\times$6.9 & & 5.5  $\pm$0.3 & 12.4   $\pm$1.9 & ...~~~~       & ...~~~~         & ...~~~~~          \\ 
II   &44.2   &2.0$\times$6.2 & & 13.2 $\pm$0.3 & 634.4  $\pm$2.5 & 4.9  $\pm$0.6 & 32.4 $\pm$0.5   & 0.0~~~~ \\ 

\hline
\end{tabular}

$^{\rm a}$distance from region {\it c} in arcsec. Negative and positive values 
correspond to regions located respectively to the southwest and northeast from 
region {\it c}.  \\
$^{\rm b}$aperture $x$ $\times$ $y$ where $x$ is the slit width and $y$ the size along the slit in arcsec. \\
$^{\rm c}$in units 10$^{-16}$\ erg\ s$^{-1}$cm$^{-2}$. \\
$^{\rm d}$in \AA. \\
\end{table*}


\begin{table*}[tbh]
\caption{Equivalent widths of   
          H$\gamma$ and H$\delta$ absorption lines in LSB regions.}
\label{t:abshghd}
\begin{tabular}{lccccccc} \hline \hline
Region & Distance$^{\rm a}$& Aperture$^{\rm b}$
& {} &$EW$(H$\delta$)$^{\rm c,d}$ &$EW$(H$\gamma$)$^{\rm c,d}$    
&$EW$(H$\delta$)$^{\rm c,e}_{B\&A}$  &$EW$(H$\gamma$)$^{\rm c,e}_{B\&A}$  \\ 
\hline

{\it a}  &--12.4~~  &2.0$\times$7.6 && 9.0:$\pm$0.6 &      ...     & 9.8:$\pm$0.6 &  ...    \\
{\it b}  & --5.8\,  &2.0$\times$4.1 && 8.6 $\pm$0.4 & 6.6 $\pm$0.5 & 8.9 $\pm$0.4 & 6.7$\pm$0.4      \\  
{\it d}  & ~\,8.0   &2.0$\times$5.5 && 7.3 $\pm$0.5 & 5.3 $\pm$0.5 & 7.8 $\pm$0.5 & 5.0$\pm$0.5      \\
{\it f}  & 23.9  &2.0$\times$4.1 && 6.2 $\pm$0.7 & 5.1 $\pm$0.9 & 6.6 $\pm$0.7 & 5.1$\pm$0.7      \\ 
{\it g}  & 28.4  &2.0$\times$4.8 && 7.2 $\pm$0.4 & 4.8 $\pm$0.9 & 7.8 $\pm$0.4 & 5.1$\pm$0.4      \\  
{\it h}  & 34.2  &2.0$\times$6.9 && 7.4 $\pm$0.5 & 5.3 $\pm$0.5 & 7.6 $\pm$0.5 & 7.4$\pm$0.5      \\
\hline
\end{tabular}

$^{\rm a}$distance from region {\it c} in arcsec. Negative and positive values 
correspond to regions located respectively to southwest and northeast from 
region {\it c}.  \\
$^{\rm b}$aperture $x$ $\times$ $y$ where $x$ is the slit width and $y$ the size along the slit in arcsec. \\
$^{\rm c}$in \AA.  \\
$^{\rm d}$equivalent widths are measured using Gaussian fitting of the lines. \\
$^{\rm e}$equivalent widths are obtained within the same wavelength intervals 
$\lambda$$_0$ = 4318 -- 4364\AA\ and 4082 -- 4124\AA, 
respectively, for H$\gamma$ and H$\delta$ as those used by Bica \& Alloin 
(\cite{Bica86}). \\ 
\end{table*}

  \begin{figure*}
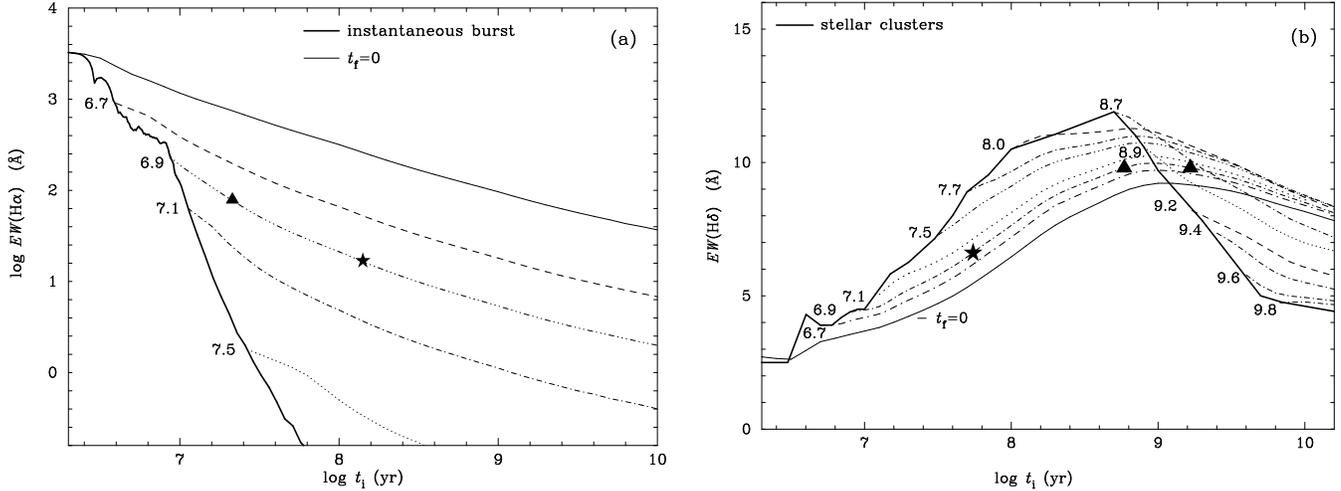

\vspace{1.cm}
    \hspace*{-0.0cm}\psfig{figure=3329.f6a.ps,angle=-90,height=6.5cm,clip=}
    \hspace*{0.4cm}\psfig{figure=3329.f6b.ps,angle=-90,height=6.5cm,clip=}
   \caption[]{({\bf a}) Temporal evolution of the equivalent width 
    of the H$\alpha$ emission line  
    for an instantaneous burst (thick solid line)
    and a heavy element mass fraction 
    $Z$ = $Z_\odot$/20 calculated with the PEGASE.2 code
    (Fioc \& Rocca-Volmerange \cite{F97}).
    Other different lines show model predictions for
    continuous star formation with constant SFR, starting at the time $t_{\rm i}$
    ago, defined by the abscissa,
    and stopping at $t_{\rm f}$, with  $t_{\rm f}$ = 0 (thin solid line)
    and $t_{\rm f}$ changing from 5 Myr to 30 Myr ago. Each line is labeled by 
    the respective log $t_{\rm f}$ ($t_{\rm f}$ in yr). 
    The positions of the observed $EW$(H$\alpha$)  
    on the modeled curve
    are shown for region {\it a} by triangle and for region {\it f}
    by star for the case of continuous star formation with 
    $t_{\rm f}$ = 8 Myr.
      ({\bf b}) Equivalent width of the H$\delta$
    absorption line vs age for stellar clusters (Bica \& Alloin 
    \cite{Bica86}) (thick solid line).
    The  predicted $EW$(H$\delta$) are also shown for continuous star 
    formation with constant SFR starting at time $t_{\rm i}$ ago, defined by the 
    abscissa, and stopping at $t_{\rm f}$, with  $t_{\rm f}$ = 0 
    (thin solid line) 
    and $t_{\rm f}$ from 5 Myr to 6.3 Gyr. Each line is labeled by 
    the respective log $t_{\rm f}$ ($t_{\rm f}$ in yr).  
    The positions of the observed $EW$(H$\delta$) 
    on the modeled curve for region {\it a}
    are shown by triangles and 
    the position of the observed $EW$(H$\delta$)
    for region {\it f} is shown
    by star for the case of continuous star formation with 
    $t_{\rm f}$ = 8 Myr. The equivalent widths  
    have been  measured in the same wavelength intervals as those 
    used by 
    Bica \& Alloin (\cite{Bica86}).
    }    
\label{fig:3}
\end{figure*}

The dependence of the H$\alpha$ and H$\beta$ 
emission line equivalent widths 
on age for a heavy element mass fraction $Z_\odot$/20
has been calculated for two limiting star formation histories,
the case of an instantaneous burst, and that of a continuous constant
star formation. The calculations are done with 
the galactic evolution code PEGASE.2
(Fioc \& Rocca-Volmerange \cite{F97}). 
The time evolution of H$\alpha$ in the case of an
instantaneous burst is shown in Fig. \ref{fig:3}a by a thick solid line. 
We next consider the case of continuous star 
formation with a constant star formation rate (SFR).
The temporal evolution of the equivalent widths of the H$\beta$ and H$\alpha$ 
emission lines is calculated using the model equivalent widths of hydrogen 
emission lines and SEDs for an instantaneous burst 
(Fioc \& Rocca-Volmerange \cite{F97}).
The temporal dependences of the H$\alpha$ emission line equivalent
width are shown in Fig.~\ref{fig:3}a by different thin lines 
for continuous star
formation starting at time $t_{\rm i}$, as defined by the abscissa value, and 
stopping at $t_{\rm f}$ (the curves in Fig.~\ref{fig:3}a are labeled by
log $t_{\rm f}$). Time is zero now and increases to the past. 
 
To compare theoretical predictions with observed data we consider 
in the following regions {\it a} and {\it f}
only, as representative of the whole LSB component. 
The positions of the measured $EW$(H$\alpha$) 
on the model curve for  continuous star formation 
with $t_{\rm f}$ = 8 Myr
are shown in Fig.~\ref{fig:3}a by a triangle for region {\it a}  and by a star
for region~ {\it f}.
 In this case the data are consistent with  star
formation starting not earlier than $t_{\rm i}$ $\sim$ 20 Myr ago for region 
{\it a} and $\sim$ 200 Myr ago for region {\it f}. If instead 
continuous star formation stopping at $t_{\rm f}$ = 0 is considered, 
then the observed $EW$(H$\beta$) and 
$EW$(H$\alpha$) are consistent with models in which star formation
starts at time $t_{\rm i}$ $\sim$ 1 -- 4 Gyr ago for region {\it a} and
$t_{\rm i}$ $\gg$ 10 Gyr ago for region {\it f}. Hence, we conclude that
for region {\it f}, models in which stars are continuously forming 
until now ($t_{\rm f}$ = 0) with a constant star formation rate are 
inconsistent with the observed $EW$(H$\alpha$) and $EW$(H$\beta$).

\subsection{Age from Balmer stellar absorption lines\label{under_2}}

Hydrogen Balmer absorption lines are detected along the slit in a large part of
HS 1442+4250. This allows us to estimate the age of  stellar
populations by  comparing  the observed hydrogen absorption
line equivalent widths with theoretical ones. 
Gonz\'alez Delgado et al. (\cite{GonLeith99b}) have calculated the
temporal dependence of the Balmer absorption lines
considering both instantaneous and continuous star formation.
Their models for an instantaneous burst predict a steady increase 
of the equivalent widths with time for ages ranging from 1 Myr to 1 Gyr,
reaching maximum values $EW$(H$\gamma$) $\sim$ 12\AA\ and 
$EW$(H$\delta$) $\sim$ 15\AA\ for a  metallicity $Z$ = $Z_\odot$/20 
(Fig.~\ref{fig:3}b).
For solar metallicity, the predicted maximum $EW$s are larger 
by $\sim$ 1\AA. However, Gonz\'alez Delgado et al. (\cite{GonLeith99b}) do not extend
calculations to ages $>$ 1 Gyr, when the equivalent widths of the absorption 
lines decrease with age (Bica \& Alloin \cite{Bica86}).

We compared the model predictions of the H$\delta$ absorption line
equivalent width for an instantaneous burst by Gonz\'alez Delgado et al. 
(\cite{GonLeith99b}) (Fig. 9 in Guseva et al. \cite{Guseva2003a})
with the empirical calibration by Bica $\&$ Alloin (\cite{Bica86})
(Fig.~\ref{fig:3}b)   and with available 
observational data for a large sample of different objects: open and globular 
stellar clusters (Bica \& Alloin \cite{Bica86}, \cite{Bica86a}), nuclei of 
normal elliptical and spiral galaxies (Bica \& Alloin \cite{Bica87}; Bica 
\cite{Bica88}; Schmidt et al. \cite{Schmidt89}, \cite{Schmidt95}; 
Saraiva et al. \cite{Saraiva2001}) with different ages and 
metallicities. No stellar clusters and galaxies with $EW$(H$\gamma$)
$>$ 9.6\AA\ and $EW$(H$\delta$)$>$ 12\AA\ were found in the samples considered. Hence, the
models by Gonz\'alez Delgado et al. (\cite{GonLeith99b}) at ages $\sim$ 1 Gyr
apparently overestimate the equivalent widths of the absorption lines.

Therefore we only use an empirical calibration of Balmer absorption line 
equivalent widths versus age by Bica \& Alloin (\cite{Bica86}) for ages 
ranging between 1 Myr and 16.5 Gyr. This calibration is based on
integrated spectra of 63 star clusters with known ages, metallicities
and reddenings. 
Later, a calibration of the H$\gamma$ and H$\delta$ absorption line equivalent
widths was derived by Schmidt et al. (1995) to apply to 
starburst events in dwarf galaxies.
The dependence of the equivalent width of the  
H$\delta$ absorption line on age for an instantaneous
burst from the data of Bica \& Alloin (\cite{Bica86}) 
is shown in Fig.~\ref{fig:3}b by thick solid line. 

We also consider the case of continuous star formation. 
For this we assume that stars are
forming with a constant star formation rate starting at time $t_{\rm i}$ 
and stopping at $t_{\rm f}$. We use the equivalent widths of hydrogen 
absorption lines 
from the empirical calibration by Bica \& Alloin (\cite{Bica86}) 
and SEDs by Fioc \& Rocca-Volmerange (\cite{F97}) for instantaneous bursts to 
calculate the temporal evolution of the equivalent widths of hydrogen 
absorption lines for continuous star formation.

  The results are given in Fig.~\ref{fig:3}b. 
The temporal dependences of the equivalent
width of the H$\delta$  absorption line 
are shown by different thin lines 
for continuous star formation starting at $t_{\rm i}$, as 
defined by the abscissa value, and stopping at $t_{\rm f}$ (the curves 
in Fig.~\ref{fig:3}b are labeled by log $t_{\rm f}$).
The equivalent width of the H$\delta$ absorption line 
in the spectrum of the stellar population formed between $t_{\rm i}$ and 
$t_{\rm f}$ corresponds in Fig. \ref{fig:3}b to the $EW$ at time 
$t_{\rm i}$.  

Stellar hydrogen absorption lines have been detected in all LSB regions.
In all regions except for region {\it a}, we 
measure  the equivalent widths of the H$\delta$ 
and H$\gamma$ absorption lines.
The H$\gamma$ absorption line in region {\it a} has not been 
used because of the strong contamination by nebular emission.
The contribution of the nebular H$\gamma$ and H$\delta$ emission lines 
to the corresponding absorption lines in the spectra of 
regions {\it a}, {\it b}, {\it d}, 
{\it f}, {\it g} and {\it h} has been removed using the IRAF
software package. For the northeastern regions {\it f}, {\it g}, and {\it h}, 
the contamination of the H$\delta$ and H$\gamma$ absorption lines by nebular 
emission is small, resulting in a correction which is
within the errors of the $EW$s.

Table~\ref{t:abshghd} lists the equivalent widths with errors of 
the H$\gamma$ and H$\delta$ absorption lines. The equivalent
widths shown in cols. (4) and (5) are measured using Gaussian fitting of the lines,
while the $EW$s in cols. (6) and (7) marked with the subscript ``B\&A'' are obtained
within the same wavelength  intervals $\lambda$$_0$ = 4318 -- 4364\AA\ and 4082 -- 4124\AA, 
respectively, for H$\gamma$ and H$\delta$ as those used by Bica \& Alloin 
(\cite{Bica86}). 

The positions of the observed $EW$(H$\delta$)  
on the model curve for continuous star 
formation with $t_{\rm f}$ = 8 Myr
are shown in Fig.~\ref{fig:3}b by  triangles for region {\it a}  
and by a star for region {\it f}.
The observed $EW$(H$\delta$) and $EW$(H$\gamma$) are consistent 
with those of stellar populations forming continuously from $\sim$ 40--250 
Myr ago to 8 Myr ago if the model calculations by Gonz\'alez Delg\'ado et al. 
(\cite{GonLeith99b}) are used.
These models are shown in Fig. 9 of Guseva et al. (\cite{Guseva2003a}).
In the case of the empirical calibration by Bica \&
Alloin (\cite{Bica86}) the observed $EW$(H$\delta$) 
(Fig.~\ref{fig:3}b) and $EW$(H$\gamma$) are 
best reproduced by a stellar population forming continuously with a constant
star formation rate starting at $t_{\rm i}$ $\sim$ 50 Myr for region
{\it f} and $\sim$ 1 Gyr for 
region {\it a} and stopping at $t_{\rm f}$ = 8 Myr.

Comparison of Figs. \ref{fig:3}a and \ref{fig:3}b shows that the ages
of the oldest stars contributing to the light of region {\it f} 
(stars) derived from the 
emission and absorption line equivalent widths are consistent and lie in the 
narrow interval between 50 -- 150 Myr. This implies that continuous star formation
stopping at $t_{\rm f}$ = 8 Myr is a reasonable scenario for region {\it f}.
However, the ages of the stellar population in region {\it a} derived from
Figs. \ref{fig:3}a and \ref{fig:3}b (triangles) are not in 
agreement. The age of the oldest stars contributing to the light of this
region as derived from the emission lines is $\sim$ 25 Myr, 
while the age derived from the absorption lines
using the Gonz\'alez Delgado et al. (\cite{GonLeith99b}) and Bica \& Alloin (\cite{Bica86})
calibrations is $\sim$ 300 Myr and $\sim$1 -- 2 Gyr, respectively. 
Evidently,
continuous star formation stopping at $t_{\rm f}$ = 8 Myr does not reproduce
the observed properties of region {\it a}.
A better agreement
is achieved if star formation in this region continues until now (i.e.
$t_{\rm f}$ = 0, thin solid lines in Fig. \ref{fig:3}a and 
\ref{fig:3}b). 
Then, 
within the errors, the observed data gives  ages  
$\sim$500 Myr (using Gonz\'alez Delg\'ado et al., \cite{GonLeith99b}) and
$\sim$1 Gyr (Fig.~\ref{fig:3}b) for the oldest stars, which 
is in better agreement with
the age estimation $\sim$2 -- 3 Gyr obtained  from the 
equivalent widths of emission lines.
Note, that choosing $t_{\rm f}$ = 0 is a limiting case which gives a maximum
age of the old stellar population in a scenario with a constant star formation rate.

\subsection{Age from the spectral energy distribution \label{SED}}

We next consider the age of the stellar population derived from the comparison
of the observed and theoretical SEDs.
The shape of the observed spectrum reflects not only the properties of the
stellar population but also reddening effects.
Therefore  only a combination of the SED method 
with those discussed in Sect. \ref{under_1} and \ref{under_2} allows us to 
simultaneously estimate age and interstellar extinction for 
the LSB regions in HS 1442+4250.

To fit the observed SEDs, we use the galactic evolution code PEGASE.2 (Fioc \& 
Rocca-Volmerange \cite{F97}) to produce a grid of theoretical SEDs
for an instantaneous burst of star formation with ages ranging between  
0 and 10 Gyr, and a heavy element mass fraction  $Z$ = $Z_\odot$/20.
An initial mass function with a Salpeter 
slope ($\alpha$ = --2.35), and upper and lower mass limits of 120 
$M_\odot$ and 0.1 $M_\odot$ are adopted. 
The contribution of gaseous emission to the total emission of regions {\it a}, 
{\it b}, {\it d}, {\it f}, {\it g} and {\it h} is small and has not been taken into 
account in the SED calculations.

  \begin{figure}[hbtp]
    \psfig{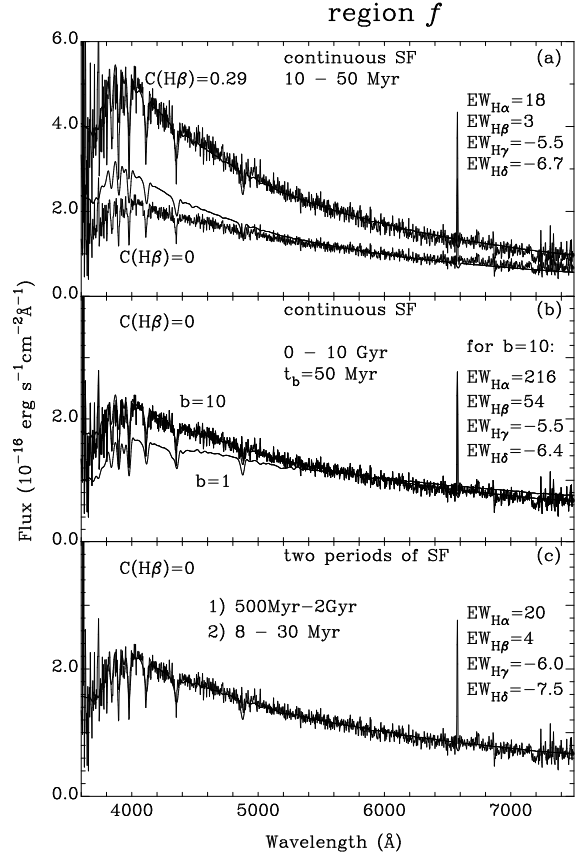}
   \caption[]{Spectrum of region {\it f} {\bf (a) -- (c)}  
 on which synthetic SEDs are 
superposed. They are calculated for stellar populations formed continuously 
within different time intervals.
The predicted equivalent widths of hydrogen emission and absorption lines
are shown for each model.
{\bf (a)} Synthetic SED corresponds to a stellar population 
formed continuously between 10 and 50 Myr ago with a constant SFR.
The lower SED is superposed on the spectrum uncorrected for extinction
and the upper SED is superposed on the spectrum corrected for extinction with
$C$(H$\beta$) = 0.29.
{\bf (b)} Synthetic SED labeled $b$ $\equiv$ 
SFR($t$ $\leq$ $t_{\rm b}$)/SFR($t$ $>$ $t_{\rm b}$) = 1
with $t_{\rm b}$ = 
50 Myr, corresponding to a stellar population formed continuously between 0 and 10 Gyr
with a constant SFR. The synthetic spectrum labeled 
$b$ = 10 corresponds to a stellar population formed continuously 
between 0 and 10 Gyr with a SFR enhanced by 10 times during  $t$ $\leq$ 
$t_{\rm b}$. {\bf (c)} Synthetic SED, calculated for star formation with 
a constant SFR occurring in two  periods: 1) 500 Myr -- 2 Gyr ago
and 2) 8 -- 30 Myr ago. The extinction coefficient is $C$(H$\beta$) = 0 for 
{\bf (b)} and {\bf (c)}.  
    }    
\label{fig:spfit_1}
\end{figure}

 A reliable determination of reddening can only be made for the
two brightest H {\sc ii} regions {\it c} and {\it e} because
of their many strong 
hydrogen emission lines (Table~\ref{t:Intens}). In other regions, only
H$\alpha$ and H$\beta$ emission lines are present. The latter line is very 
weak in regions {\it f}, {\it g} and {\it h}. The extinction coefficient 
$C$(H$\beta$) obtained from the Balmer decrement in the different regions ranges 
from 0 to 0.15.  $C$(H$\beta$) is 0.11 in the brightest region 
{\it c}.

We consider only the properties
of region {\it f}, taken to be representative of the LSB region.
We showed in Sect. \ref{under_1} and \ref{under_2} that the observed 
equivalent widths of the hydrogen emission and absorption lines 
in this region can be 
reproduced by a stellar population formed continuously with a constant SFR, 
starting at $t_{\rm i}$ $\sim$ 100 Myr and stopping at $t_{\rm f}$ = 8 Myr. 
However, the observed SED of region {\it f} cannot be reproduced by a
synthetic SED with such a young stellar population (lower spectra in 
Fig~\ref{fig:spfit_1}a) without assuming a non-negligible
 reddening. In particular, the observed SED can be fitted by 
a synthetic SED of a stellar population formed continuously between 
10 and 50 Myr with a constant SFR, if there is an extinction 
$C$(H$\beta$) = 0.29 (upper spectra in Fig.~\ref{fig:spfit_1}a).

Alternatively, the observed SED of region {\it f} 
at $\lambda$ $\ga$ 3900\AA\ can also be well fitted by a 
synthetic SED of a stellar population formed continuously  between 0 and 10 Gyr, 
if the SFR during the last 50 Myr was enhanced by a factor $b$ of 10 and no 
reddening is present (SED labeled ``b=10'' in Fig.~\ref{fig:spfit_1}b). 
However, the fit is not satisfactory at shorter wavelengths. Additionally, this model 
does not reproduce the observed $EW$(H$\alpha$) and $EW$(H$\beta$).  

We also consider models with star formation occurring in two episodes 
separated by a quiescent period.
A model with star formation occurring with a constant SFR during
(1) 1 -- 10 Gyr and (2) 10 -- 50 Myr with comparable masses of 
stellar populations formed in each period (not shown in Fig.~\ref{fig:spfit_1})
 is able to reproduce the observed
$EW$s of the hydrogen lines ($EW$(H$\alpha$) = 11\AA, $EW$(H$\beta$) = 2\AA, 
$EW$(H$\gamma$) = --5.7\AA\ and $EW$(H$\delta$) = --6.7\AA).
However, in that case the observed SED is again not well fitted at $\lambda$ $<$ 3900\AA.
All the observational data for region {\it f} are best fitted by a model in
which the stellar population formed continuously with a constant SFR in two 
periods 1) 0.5 -- 2 Gyr and 2) 8 -- 30 Myr ago and no reddening is present 
(Fig.~\ref{fig:spfit_1}c). If some extinction is present then 
ages will be smaller.

In summary, we conclude that only young and intermediate-age 
stellar populations can reproduce the observed properties of region {\it f}.

\subsection{Age from the colour distribution \label{coldist}}

We derived $V$ and $I$ surface brightness and colour distributions  for
the regions covered by the spectroscopic observations 
and compared them with predictions from our
population synthesis modeling. The results of this comparison are shown in
Fig.~\ref{f2}c. The predicted colours obtained from convolving the theoretical 
SEDs with the appropriate filter bandpasses are shown by different symbols.
The transmission curves for the Johnson $V$ and Cousins $I$ bands are taken
from Bessell (\cite{B90}). The zero points are from Bessell et al.
(\cite{B98}). 

The contribution of the gaseous emission to the total brightness is 
small in the LSB component. Therefore, we do not take it into account.
The colours of regions {\it a} and {\it b} are 
well fitted by those of a stellar
population formed continuously with a constant SFR in two periods of star
formation: (1) 100 -- 400 Myr and (2) 4 -- 15 Myr ago.
The  colours of the northeastern part of the galaxy 
excluding the brightest region {\it c} are best fitted by those of a stellar
population forming continuously  with a constant SFR in two 
periods: (1) 0.5 -- 2 Gyr and (2) 8 -- 30 Myr ago. In all cases we adopt
$C$(H$\beta$)=0, which gives upper limits to the ages. 

Since the contribution of  ionized gas emission to the total light of 
the brightest H {\sc ii} region (region {\it c}) is high, the theoretical SED 
for this region has been constructed using a 4 Myr stellar population SED for a 
heavy element mass fraction $Z$ = $Z_\odot$/20, and adding
the gaseous continuum SED and observed emission lines 
(see Guseva et al. 2001 for details). The 
predicted $(V-I)$ colour of region {\it c} is reddened adopting 
$C$(H$\beta$) = 0.11 (Table~\ref{t:Intens}). The observed colour $(V-I)$ 
$\sim$ --0.5 mag of this region is bluer than that of a 4 Myr stellar population
(labeled by an asterisk in Fig.~\ref{f2}c) implying a large contribution of 
ionized gas emission (labeled by a triangle in Fig.~\ref{f2}c).

In Fig.~\ref{f2}c we show with filled circles the modeled colours of all regions in 
the LSB component of the galaxy and the total colour of region {\it c}.
The agreement between the observed and synthetic $(V-I)$ colours is very good.
  The reddest colours ($V-I$) $\sim$ 0.63 mag in \sbs\ are found in the outer parts 
of the galaxy at $\mu_V$ $\ge$ 24 \sbu\ (Fig.~\ref{f4}). No 
spectroscopic data are available for these regions. Therefore, we can 
only use photometric data to estimate 
the age of the stellar population in those regions. 
Again, the maximum age depends on 
the adopted star formation history. If an instantaneous burst model is assumed
then the age of the population with ($V-I$) = 0.63 mag is $\sim$1 Gyr.
In the case of continuous star formation the same colour can be explained 
by a stellar population formed with a constant star formation rate between 
0 and 5 Gyr.  
We note, however, that the age estimates for the outer part of \sbs\ are uncertain
due to large errors in the ($V-I$) colour and likely contamination of
the LSB light by red background/foreground objects.

 \section{Conclusions \label{conc}}

We present  
a detailed photometric and spectroscopic study
of the low-metallicity dwarf irregular galaxy HS 1442+4250.
Broad-band $V$ and $I$ images 
and  spectra in the optical range have been obtained 
with the 2.1m and 4m Kitt Peak telescopes respectively.
The main conclusions of this study can be summarized as follows:

\begin{enumerate}

\item HS 1442+4250 is a low-metallicity nearby ($D$ = 12.4 Mpc) 
dwarf irregular galaxy
with a chain of H {\sc ii} regions arranged along an elongated 
low-surface-brightness (LSB) component. 
The LSB component is well fitted by an exponential profile with
a scale length $\alpha$ $\ga$ 270 pc in its outer part, 
flattening for small radii ($R^*$ $\la$ 2$\alpha$).
The observed $(V-I)$ colour of the brightest H {\sc ii} region  is very blue 
$\sim$ --0.5 mag due to the combined effect of a
young stellar population and ionized gas emission. The colour of the 
LSB component is constant and equal to $\sim$ 0.4 mag in a large part 
of the galaxy.

\item The oxygen abundance for the brightest H {\sc ii} region {\it c} is 
12 + log(O/H) = 7.63 ($Z$ = $Z_\odot$/19). 
The neon-to-oxygen abundance ratio for this region, log Ne/O = --0.75,  is 
in good agreement with the mean ratio derived from  previous studies of
dwarf galaxies (e.g., Izotov \& Thuan \cite{IT99}).
The nitrogen-to-oxygen abundance ratio log N/O = --1.44 is $\sim$ 0.15 dex  
larger than the N/O ratios obtained
by Thuan et al. (\cite{til95}) and Izotov \& Thuan (\cite{IT99}) for the most
metal-deficient blue compact dwarf (BCD) galaxies with $Z$ $\la$ $Z_\odot$/20. 

\item The $^4$He mass fraction $Y$ = 0.243 $\pm$ 0.008 
derived for the brightest H {\sc ii} region is 
in good agreement with the primordial $^4$He mass fraction $Y_{\rm p}$ = 0.244 -- 0.245
derived by Izotov \& Thuan (\cite{IT98}) and Izotov et al. (\cite{ICFGGT99}).
   
\item We use four methods to estimate the age of the stellar 
population in the LSB component of HS 1442+4250.
Different histories of star formation are considered.
The spectroscopic data such as the equivalent widths of 
the hydrogen H$\alpha$ and H$\beta$ emission lines, and
of the hydrogen H$\gamma$ and H$\delta$ absorption lines,
the spectral energy distributions and the photometric colours are 
reproduced quite well by models with only young and intermediate-age 
stellar populations.
It is unlikely that a significant old stellar population contributes
much to the observed light in the inner regions.
 However, in the outermost regions of the LSB component, with 
($V-I$) $\sim$0.63,  the presence of  older populations is not excluded.
For these regions only photometric data are available.
The observed colour  can be explained
by a stellar population continuously forming between 0 and 5 Gyr with
a constant SFR. However, because of large photometric errors 
and possible confusion with background sources,
those age estimates are uncertain.

\end{enumerate}

\begin{acknowledgements}
N.G.G. and Y.I.I. acknowledge DFG grant 
436 UKR 17/2/02 and Y.I.I. is grateful for the Gauss professorship
of the G\"{o}ttingen Academy of Sciences.
They are also grateful for  Swiss
SCOPE 7UKPJ62178
grant and for hospitality at G\"ottingen Observatory. 
Y.I.I. and T.X.T. acknowledge 
partial financial support through NSF grant AST-02-05785.
Research by P.P. and K.J.F. has been supported by the
Deutsches Zentrum f\"{u}r Luft-- und Raumfahrt e.V. (DLR) under
grant 50\ OR\  9907\ 7.  K.G.N. thanks support from the Deutsche
Forschungsgemeinschaft (DFG) grants FR 325/50-1 and FR 325/50-2.
\end{acknowledgements} 

{}
\end{document}